# Deep Learning Based Resource Allocation for Full-duplex Device-to-Device Communication

Xinxin Zhang [1], Lei Gao[2*]

1. School of Information and Communication Engineering, Beijing University of Posts and Telecommunications, Beijing, China
2. 6G Research Center, China Telecom Research Institute, Beijing, China
zhangx2@bupt.edu.cn, gaol8@chinatelecom.cn
Corresponding Author: Lei Gao    Email: gaol8@chinatelecom.cn

*Abstract*—Device-to-device (D2D) technology is one of the key research areas in 5G/6G networks, and full-duplex (FD) D2D will further enhance its spectral efficiency (SE). In recent years, deep learning approaches have shown remarkable performance in D2D resource allocation tasks. However, most schemes only model the channel state information (CSI) as an independent feature, neglecting the spatial relationships among multiple channels and users within the scenario. In this paper, we first design an objective function for FD D2D communication resource allocation, which aims to maximize the SE of D2D users while ensuring the minimal required SE of cellular users. Then, considering the complex CSI constituted by all the users in different channels as a three-dimensional vector, a centralized resource allocation model based on multi-dimensional spatial convolutional networks and attention mechanisms (SP-Conv-Att) is proposed. To alleviate the burden of base station, we develop two distributed models, Dist-Att and Dist-Att-Conv, to facilitate users to perform channel and power allocation locally, based on attention and multi-user convolutional networks respectively. Numerical results demonstrate that our models outperform traditional schemes and recent deep neural network models, significantly approximating the optimal solution computed by exhaustive algorithm with extremely low latency.

## I. Introduction

Device-to-device (D2D) is one of the potential technologies in the process of 5G/6G network construction. Within the same network, D2D users (DUEs) are capable of sharing the radio resources of cellular users (CUEs), which will improve the performance of the whole system in terms of spectrum utilization and network capacity [1]. In this scenario, signal transmission within the same channel will cause mutual interference. When time is not considered, exhaustive algorithm will achieve the optimal transmission scheme. However, excessive communication delay is a serious problem. To achieve fast and intelligent interference control, existing literature has optimized resource allocation tasks by setting different objectives, such as improving system sum-rate [2], reducing interference power experienced by users [3], maximizing energy efficiency (EE) of the system [4], reducing strategy complexity [5], and enhancing distributed performance [6]. These optimizations mainly focus on the half-duplex (HD) scenarios, where the maximum communication capacity of the system is relatively lower. Meanwhile, full-duplex (FD), as another data communication mode, is considered to have great potential to improve the spectral efficiency (SE) and reliability of the whole system [7].

Theoretically, the SE in HD mode is expected to double in FD mode. However, there are two major issues that need to be addressed. 1) In FD mode, D2D users in a pair will transmit and receive signals simultaneously, resulting in a larger scale of interference; 2) Unable to completely eliminate self-interference, which will limit the SE of users. Research has found that some approaches have optimized the D2D resource allocation under FD through classic strategies such as function transformation [8], alternating optimization [9], bipartite matching [10], and game theory [11]. However, most algorithms are not sufficiently novel and intelligent.

With the growing research interest in artificial intelligence (AI), data-driven deep learning (DL) models have successfully addressed resource allocation in various scenarios such as vehicular networks [12]], 5G broadband services [13], and video communications [14]. Similarly, the great potential of DL technology has been highlighted with its increasing application in the field of D2D communication. [15] proposed a double deep Q-network based resource matching algorithm for multiuser communication to improve the rate and EE of the single user with better convergence speed. [16] combined deep neural networks (DNN) and heuristic algorithms to maximize the total rate of DUEs under a given QoS constraint. [17] jointly designed a power control and location scheduling scheme based on a graph neural network to support multi-UAV D2D communication, maximizing the total downlink user transmission rate with low-complexity. In order to alleviate the burden on the base station (BS), [18] implemented power and channel configuration for D2D users by designing a distributed model based on DNN. [19] developed a distributed resource allocation algorithm combining deep Q network and unsupervised learning. For FD scenarios, [20] constructed a neural network (NN) to optimize the EE for

D2D systems and experiments showed that the prediction of the model will be close to the optimal solution of the exhaustive algorithm. [7] proposed a DNN-based transmit power allocation approach that automatically determine the optimal power for users in the same spectrum.

Despite the fact that AI techniques effectively facilitate D2D communication, most of the schemes only characterize the channel state information (CSI) in the scenarios as an independent dimension, which ignores the three-dimensional spatial information consisting of resources such as users and channels. Further investigation reveals that designing DL models in a distributed manner is rarely considered in FD D2D scenarios, however, the distribution will significantly reduce the burden at the BS, which is very advantageous in some scenarios. Meanwhile, AI, as the core technology of endogenous intelligence for 6G, needs to be further explored for its possibilities in network resource allocation.

Motivated by the aforementioned challenges, we propose several DL-based FD D2D resource allocation models. The contributions of this paper are outlined below:

1) An objective function is given to simultaneously solve the channel allocation and power level prediction tasks in a FD scenario, maximizing the SE of DUEs while ensuring the SE of CUEs above a threshold;

2) Using the BS as a computational center, a model incorporating the dimensions of channel and user is proposed based on spatial convolutional neural network and attention mechanism (SP-Conv-Att), which are used to approximate the optimal solution of the proposed objective function (NP-hard problem);

3) Based on attention and multi-user convolutional network, we develop two distributed models, Dist-Att and Dist-Att-conv, which trade off slightly degraded performance for reduced burden at the BS;

4) Simulations reveal that our models remarkably approach the results of the optimal scheme while reducing the computational time by a considerable margin.

The remaining structure of this paper is organized as follows. Chapter II gives the definition of the FD D2D communication system model and its optimization problem. Chapter III proposes several resource allocation schemes based on DL. Chapter IV designs an experiment to evaluate our schemes. Chapter V presents the conclusions.

## II. SYSTEM MODEL AND PROBLEM STATEMENT

In this paper, we consider resource allocation for the uplink link, where DUEs are FD communication and CUEs are HD communication. The scenario diagram is shown in Fig. 1. The number of D2D-pair is $N$ ($\hat{N} \in \{1,...,N\}$, $\hat{N}$ is the entire set) and the number of CUEs is $K$ ($\hat{K} \in \{1,...,K\}$, $\hat{K}$ is the entire set). We assume that all users are equipped with a single antenna and are randomly distributed in an area of $X*Y$. For CUEs, each cellular user occupies one channel. In each D2D-pair, two users occupy the same channel and are labeled as $a$ and $b$. The BS is located in the middle of the cell. In this network, the DUEs are allowed to reuse the channels of the CUEs to improve the SE of the system.

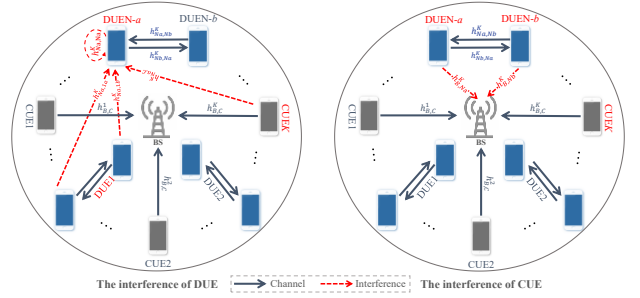

Fig. 1. FD D2D scenario (Note: not all interferences are depicted).

In Fig. 1, the left red dashed line represents the interference experienced by a D2D user, while the right red dashed line depicts the interference encountered by the BS. The gain/interference in the scenario is composed of path loss and multipath fading. To gauge the levels of gain/interference, we define some symbols. For the DUEs, $h_{ia,ib}^k$ is the effective gain generated by user $b$ of the $i$-th D2D-pair towards user $a$ of the same D2D-pair within channel $k$. For a cellular user in the $k$-th channel, $h_{B,C}^k$ is used to measure the signal gain that is transmitted to the BS. In addition, the interferences of DUEs come from three components:

- Cellular user: i.e., $h_{iu,C}^k$ denotes the interference originated by a cellular user to the $i$-th D2D-pair in channel $k$, where $u \in \{a,b\}$;
- Other D2D-pairs: i.e., $h_{iu,ju}^k$ denotes the interference generated by the $j$-th D2D-pair to the $i$-th D2D-pair, where $i$ and $j$ are not equal;
- Self-interference: i.e., $h_{iu,iu}^k$ denotes the interference produced by the $i$-th D2D-pair in channel $k$ to itself.

Compared to DUEs, BS has a single interference source, emanating solely from all D2D-pairs within the channel, i.e., $h_{B,iu}^k$ denotes the interference generated by the $i$-th D2D-pair to the BS in channel $k$. In summary, the generated gains and interferences constitute a full CSI vector ($\hat{\mathbf{H}}$) for the system. The objective function is set as

$$\mathcal{P}: \max_{f_c^{i,k}, f_p^{iu}} \xi \left[ \sum_{i \in \hat{N}} \sum_{k \in \hat{K}} \sum_{u \in \{a,b\}} \text{SE}_{iu}^k \right]$$

$$\text{s.t.} \quad C1: f_c^{i,k}(\hat{\mathbf{H}}) \in \{0,1\} \quad \forall i \in \hat{N}, \ \forall k \in \hat{K}$$

$$C2: \sum_{k \in \hat{K}} f_c^{i,k}(\hat{\mathbf{H}}) \leq 1 \quad \forall i \in \hat{N}, \ \forall k \in \hat{K}, \quad (1)$$

$$C3: f_p^{iu}(\hat{\mathbf{H}}) \in \hat{P} \quad \forall i \in \hat{N}, \ u \in \{a,b\}$$

$$C4: \text{SE}_C^k > CUE_{thr} \quad \forall k \in \hat{K}$$

which aims to maximize the SE of DUEs under the fulfillment of four constraint conditions. The combination of $C1$ and $C2$ indicates that one user occupies only one channel, where $f_c(\hat{\mathbf{H}})$ defines the channel allocation task. $C3$ represents the range for power selection, where $f_p(\hat{\mathbf{H}})$ denotes the transmit power level prediction task. In particular, with reference to 3GPP, power discrete values will outperform continuous values during deployment [18], i.e., $f_p(\hat{\mathbf{H}}) \in \hat{P} = \{P_1,...,P_L\}$, where $L$ is the maximum power level, $\hat{P}$ is the set of powers, and the power is uniformly divided into $L$ classes. $C4$ constrains the SE of CUEs to exceed a threshold $CUE_{thr}$. $\xi$ represents the expected value.

The SE for user $a$ in D2D-pair is presented as

$$SE_{ia}^k = \log_2\left(1 + \frac{h_{ia,ib}^k f_c^{i,k}(\hat{\mathbf{H}}) f_p^{ib}(\hat{\mathbf{H}})}{N_0 W + n_{ia,DUEs}^k + h_{ia,C}^k P_C + SI_{ia}^k}\right), \quad (2)$$

which is the expression resulting from the Shannon capacity theorem divided by bandwidth. Where, $N_0W$ represents Gaussian white noise, $n_{ia,DUEs}^k$ represents the interference received by the user from other DUEs:

$$n_{ia,DUEs}^k = \sum_{j \in \hat{N}(\neq i)} \sum_{u \in \{a,b\}} h_{ia,ju}^k f_c^{j,k}(\hat{\mathbf{H}}) f_p^{ju}(\hat{\mathbf{H}}), \quad (3)$$

the self-interference $SI_{ia}^k$ is formulated as

$$SI_{ia}^k = \eta_{ia} f_c^{i,k}(\hat{\mathbf{H}}) f_p^{ia}(\hat{\mathbf{H}}), \quad (4)$$

with $\eta$ is the self-interference mitigation factor and ranging from 0 to 1.

Similarly, for user $b$ in D2D-pair, the $SE_{ib}^k$, the $n_{ib,DUEs}^k$ and the $SI_{ib}^k$ are expressed as

$$SE_{ib}^k = \log_2\left(1 + \frac{h_{ib,ia}^k f_c^{i,k}(\hat{\mathbf{H}}) f_p^{ia}(\hat{\mathbf{H}})}{N_0 W + n_{ib,DUEs}^k + h_{ib,C}^k P_C + SI_{ib}^k}\right), \quad (5)$$

$$n_{ib,DUEs}^k = \sum_{j \in \hat{N}(\neq i)} \sum_{u \in \{a,b\}} h_{ib,ju}^k f_c^{j,k}(\hat{\mathbf{H}}) f_p^{ju}(\hat{\mathbf{H}}) \quad (6)$$

and

$$SI_{ib}^k = \eta_{ib} f_c^{i,k}(\hat{\mathbf{H}}) f_p^{ib}(\hat{\mathbf{H}}), \quad (7)$$

respectively.

Finally, the SE of CUEs and the interference received from DUEs are denoted as

$$SE_C^k = \log_2\left(1 + \frac{h_{B,C}^k P_C}{N_0 W + n_{B,DUEs}^k}\right) \quad (8)$$

and

$$n_{B,DUEs}^k = \sum_{j \in \hat{N}} \sum_{u \in \{a,b\}} h_{B,ju}^k f_c^{j,k}(\hat{\mathbf{H}}) f_p^{ju}(\hat{\mathbf{H}}), \quad (9)$$

respectively.

## III. DL-BASED RESOURCE ALLOCATION SCHEME

As shown in (1), the objective function to be solved belongs to a non-convex optimization problem with integer constraints, and its analytical solution cannot be obtained. To respond to these challenges, in a centralized and distributed manner, we design several models to handle the channel allocation and power level classification tasks.

### A. SP-Conv-Att Model

In the centralized approach, The BS collects all measured CSI from the receivers of the users and determines the configuration of the transmission channel and power. Then, the allocation scheme is notified to the DUEs by the BS, and data transmission begins. As multiple users can choose different channels, it will form a three-dimensional CSI feature, which can be viewed as having spatial properties. To integrate the information from both the user and channel dimensions, the SP-Conv-Att model is proposed based on a multi-dimensional spatial convolution and attention mechanism.

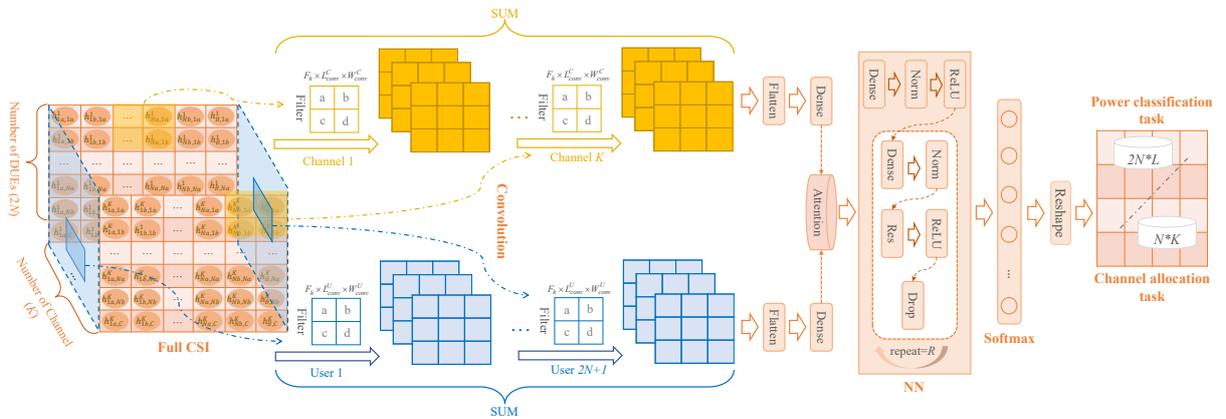

Fig. 2. SP-Conv-Att model diagram.

The overall schematic diagram is shown in Fig. 2. Firstly, the model extracts deeper features from both the channel and user dimensions based on three-dimensional convolution. After flattening into one dimension, a Dense layer is used to transform them into vectors of size $F_h$. Then, an attention mechanism is designed to assign weights to each dimension of the vector, achieving automated feature fusion and obtaining a new vector of size $F_h$. The vector will go through $R$ times NN, including Dense, Batch Normalization (Norm), ReLU, Residual (Res), and Dropout (Drop) modules. After a Softmax and Reshape operation, the probability distribution values of [$2N$, $L$] or [$N$, $K$] are output, thus completing the prediction of power levels or channels for each user.

The core of the model is the multi-user and multi-channel convolution operation. Multi-user convolution refers to performing cross-channel convolutions and summing them across multiple users, represented as

$$O_{Length}^{U} = \frac{(2N+1) + 2M_U - L_{conv}^{U}}{S_U} + 1 \quad (10)$$

and

$$O_{width}^{U} = \frac{K + 2M_U - W_{conv}^{U}}{S_U} + 1 \quad (11)$$

in the length and width dimensions, respectively. Here, $2N+1$ and $K$ represent the length and width of the input features, which correspond to the number of users and channels. $M_U$ represents the size of matrix boundary padding. $L_{Conv}^{U}$ and $W_{Conv}^{U}$ denote the length and width of the convolution kernel, respectively. $S_U$ is the stride and $O^U$ is the size of the output.

Multi-channel convolution, i.e., performing cross-user convolutions and summing them across multiple channels, has both the length and width equal to the number of users. The convolution formulas are expressed as

$$O_{Length}^{K} = \frac{(2N+1) + 2M_K - L_{conv}^{K}}{S_K} + 1 \quad (12)$$

and

$$O_{width}^{K} = \frac{(2N+1) + 2M_K - W_{conv}^{K}}{S_K} + 1, \quad (13)$$

respectively, which are computed similarly to multi-user convolution.

### B. Dist-Att Model

Although the centralized models will obtain the CSI of the entire environment and perform more accurate resource allocation, the burden at the BS is enormous, making it difficult to deploy successfully in certain situations. Therefore, we propose a distributed FD D2D resource allocation model, named the Dist-Att model, based on neural networks and attention mechanisms.

**Algorithm 1** FD D2D resource allocation algorithm based on neural networks and attention mechanisms

**Input:** $N$ DUEs, $K$ CUEs, CSI of the local user with size [$2N+1$, $K$], training rounds $R_t$.
**Initialization:** All parameters of the networks, Initial training epoch $r_t=1$.
**Repeat**
  Step 1: Users Side
    ◇ Information Extraction: NN module (with Dense, Norm, ReLU, Res, Drop, etc.) is designed to extract features from the original CSI (denoted by $\mathbf{U_{st}^{i'}}$, $i' \in \hat{I} = \{1, 2, ..., 2N+1\}$) to get the $\mathbf{U_B^{i'}}$;
    ◇ Information Transmission: Send $\mathbf{U_B^{i'}}$ to the BS.
  Step 2: BS Side
    ◇ Aggregate Information: Collecting $\mathbf{U_B^{i'}}$ from users;
    ◇ Designing Attention Mechanism: Add an adaptive weight to each CSI for each user to measure its importance and output a new message $\mathbf{U_{Att}}$, as shown in (14)-(15);
    ◇ Information Return: Dist-Att/Dist-Att-Conv
      • Dist-Att: Using only $\mathbf{U_{Att}}$, generate information for each D2D user based on NN, and distribute it, i.e. $\mathbf{U_{B\_new}^{i'}}$ ($i' \neq 2N+1$).
      • Dist-Att-Conv: $Concat(\mathbf{U_B}, \mathbf{U_{Att}})$, generate new information for each D2D user based on convolutional networks, and distribute it, i.e. $\mathbf{U_{B\_new}^{i'}}$ ($i' \neq 2N+1$).
  Step 3: DUEs Side
    ◇ Information Concatenate: $Concat(\mathbf{U_{st}^{i'}}, \mathbf{U_{B\_new}^{i'}})$;
    ◇ Resource Allocation: NN module is designed to accomplish the tasks of channel allocation and power level classification, and obtain the predicted channel $y_c$ and power $y_p$ respectively.
  Step 4: Loss Calculation and Parameter Update
    ◇ Loss of the model is calculated according to (16)-(18), where $\tilde{y}$ indicates the target value, calculated by an exhaustive algorithm;
    ◇ Update parameters based on an Adam algorithm.
**Until** $r_t > R_t$
**Output:** Channels and power levels of the DUEs.

$$\mathbf{U_{Att}} = \sum_{i' \in \hat{I}} \left( \mathbf{U_B^{i'}} \times Softmax(\mathbf{W_{Att}} \mathbf{U_B}) \right) \quad (14)$$

$$\mathbf{U_B} = Concat\left(\mathbf{U_B^1}, ..., \mathbf{U_B^{2N+1}}\right) \quad (15)$$

$$L_{total} = \sum_{i' \in \hat{I}(\neq 2N+1)} \left(L_c^{i'} + L_p^{i'}\right) \quad (16)$$

$$L_c^{i'} = \sum_{k \in \hat{K}} \tilde{y}_c^{i'k} \log(y_c^{i'k}) \quad (17)$$

$$L_p^{i'} = \sum_{l=1}^{L} \tilde{y}_p^{i'l} \log(y_p^{i'l}) \quad (18)$$

The overall schematic diagram of the model is shown in Fig. 3. It consists of three parts: Users Side, BS Side,

and DUEs Side, corresponding to the first three steps in Algorithm 1. For the Users Side, the model completes information extraction and transformation, generating $F_B$ features for each user to provide to the BS; For the BS Side, the model performs information aggregation and processing tasks. It utilizes attention to fuse the features transmitted by multiple users, and generates $F_D$ global features that need to be sent back to each D2D user based on NN modules; For the DUEs Side, users execute power level and channel classification based on local CSI and $F_D$ global features. For detailed information, please refer to Algorithm 1.

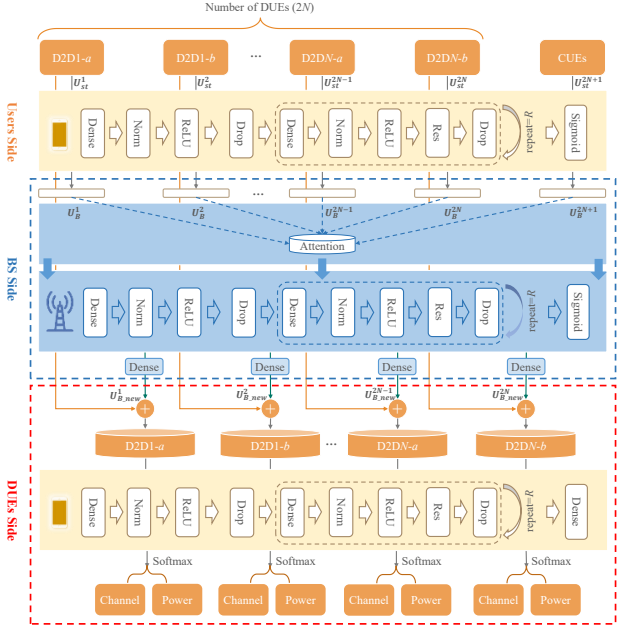

Fig. 3. Dist-Att model diagram.

## C. Dist-Att-Conv Model

From the BS Side of the distributed model, it can be observed that the design of Attention will compress the feature size $(2N+1)*F_B$ to $F_B$, greatly reducing the number of model parameters that the BS needs to handle. However, this also poses a risk of information loss. In some scenarios, when the load is sufficient to support more parameter calculations, using only the features after Attention will waste the resources of the BS. To avoid this situation, we designed a multi-user convolutional network, i.e., the Dist-Att-Conv model.

The model schematic diagram is shown in Fig. 4, where the improved part corresponds to the BS Side of the Dist-Att model, as described in Algorithm 1. The features received from multiple users are weighted and fused by the attention mechanism to obtain $\mathbf{U_{Att}}$. Subsequently, the user features $\mathbf{U_B}$ and $\mathbf{U_{Att}}$ are concatenated together, and a multi-user convolution is performed, whose convolution formulas in length and width dimensions are denoted as

$$O_{Length}^{Dist} = \frac{(2N+2) + 2M_{Dist} - L_{conv}^{Dist}}{S_{Dist}} + 1 \quad (19)$$

and

$$O_{width}^{Dist} = \frac{F_B + 2M_{Dist} - W_{conv}^{Dist}}{S_{Dist}} + 1, \quad (20)$$

respectively. The operation is similar to (10)-(11), except that the input length is $2N+2$, the width is $F_B$, and it is only a two-dimensional convolution.

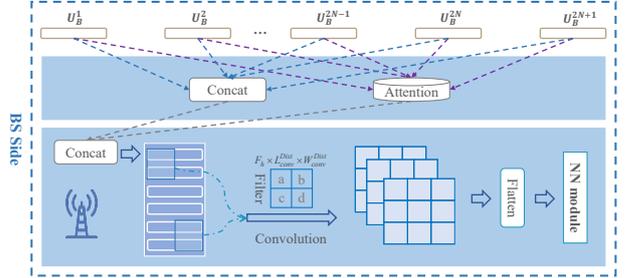

Fig. 4. Dist-Att-Conv model diagram.

## IV. EXPERIMENT

To demonstrate the efficacy of the proposed model, we utilize Python 3.7 and a GPU platform (NVIDIA TITAN Xp) to construct a scenario. The minimal SE of CUEs is set to $CUE_{thr}$=0.5b/s/Hz for experiments.

### A. Simulation and Parameters of D2D Communication

In our experiment, there are two cellular users and four D2D users (two pairs), i.e., $K=N=2$. The maximum distance between D2D users is 30m. The coefficient and exponent of the path loss model are $10^{3.453}$ and 3.8, respectively. The small-scale fading is simulated as independent identically distributed circularly symmetric complex Gaussian random variables [18]. The remaining scenario parameters are shown in Table I. Where, the self-interference cancellation coefficient $\eta$ is set to -100dB, which is relatively easy to achieve in practice [10].

TABLE I. PARAMETERS FOR FD D2D SCENARIO

| X/Y | W | $N_0$ | L | $P_0$ | $P_8$ | $\eta$ |
|---|---|---|---|---|---|---|
| 100 m | 10 MHz | -173 dBm/Hz | 8 | 0 mW | 200 mW | -100 dB |

TABLE II. PARAMETERS FOR DL MODELS

| R | $R_t$ | Learning rate | $F_B$ | $F_D$ | Hidden neurons/$F_h$ |
|---|---|---|---|---|---|
| 3 | 30~50 | 0.0001 | 10 | 32 | 256 |

For the DL models, the train set: development set (DEV): test set (TEST) = 161740: 17972: 19968, the loss of the model is calculated using a cross-entropy function, the parameters are updated based on the Adam algorithm, the training batch size is 1024, the length and width of all convolution kernels are 2, and the other parameters are presented in Table II. Where, the training round $R_t$ is set to a range value of 30~50, and its size will be adjusted to avoid overfitting and underfitting. Before training the model, its prediction target is obtained by an exhaustive algorithm, i.e., the optimal scheme: Iterating over all possible allocation schemes and outputting the optimal channel and power combination.

*B. Comparison Models*

Comparison models used in this experiment include a traditional approach ERP and a DL-based approach FC-DNN, which are described as follows:

1) ERP [21]: Equally reduced power, meaning that the transmission power of all DUEs is uniformly decreased while satisfying the conditions C1~C4 as specified in (1).

2) FC-DNN: A DNN model based on the full CSI. It treats each CSI value as independent feature and passed through modules such as Dense, Batch, ReLU, Drop, Res, etc. The authors have already demonstrated its excellent performance in D2D resource allocation task [18]. We extend it to the FD scenario.

*C. Performance of Centralized Full-duplex Resource Allocation Models*

The resource allocation accuracy of the centralized model is shown in Table III. The comparison between the DEV and the TEST indicates that the generalization performance of all models is good. Compared to FC-DNN, SP-Conv-Att outperforms in two resource allocation tasks, achieving the highest improvement of 3.4% in channel prediction task on the test set.

TABLE III. ACCURACY (%) OF MODELS

| Model | DEV | | TEST | |
|---|---|---|---|---|
| | $f_c(\hat{\mathbf{H}})$ | $f_p(\hat{\mathbf{H}})$ | $f_c(\hat{\mathbf{H}})$ | $f_p(\hat{\mathbf{H}})$ |
| FC-DNN | 85.86 | 65.74 | 86.14 | 65.64 |
| SP-Conv-Att | 89.54 | 67.82 | 89.54 | 67.79 |
| Dist-Att | 85.11 | 63.86 | 85.48 | 64.03 |
| Dist-Att-Conv | 87.56 | 66.16 | 88.18 | 66.49 |

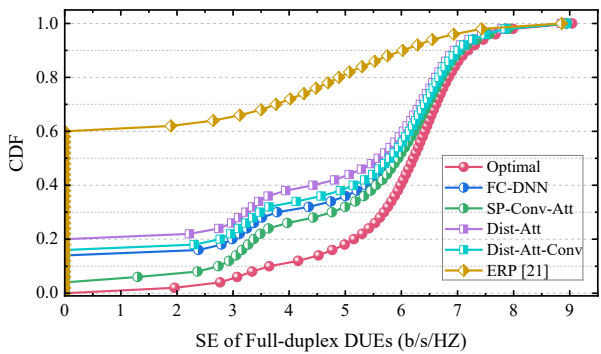

Fig. 5. CDF of SE for DUEs when CUEthr =0.5b/s/Hz.

Meanwhile, Fig. 5 shows the cumulative probability distribution function (CDF) of average SE for DUEs in the test samples. The curve of SP-Conv-Att is closer to the optimal solution, indicating that the predicted channel and power allocation scheme is generally superior. This demonstrates the effectiveness and superiority of multi-dimensional spatial convolution operations. Comparing it with the traditional ERP scheme, the DL-based model significantly reduces the gap to the optimal solution. What needs to be further explored is that ERP under continuous power will be more likely to find excellent resource allocation combinations than discrete power.

From the figure, it can also be seen that due to the high complexity of the scenario, all models have cases where SE is 0. The comparison reveals that SP-Conv-Att reduces the probability of this occurrence by approximately 10% compared to FC-DNN, further highlighting its superiority.

*D. Performance of Distributed Full-duplex Resource Allocation Models*

The accuracy of the distributed models is shown in Table III. Among them, Dist-Att-Conv has higher accuracy than Dist-Att in both channel and power prediction tasks, with a maximum difference of 2.7%. The CDF curve in Fig. 5 indicates that compared to using Attention only, Dist-Att-Conv is closer to the optimal solution of the sample and reduces the probability of SE=0 by about 4%. These results confirm the effectiveness of performing multi-user convolution operations at the BS Side.

It is worth noting that compared to FC-DNN, the Dist-Att-Conv model has higher accuracy but a worse CDF. This is because when the predicted results of the model do not perfectly match the target values, it may still calculate a suboptimal solution, and the output SE is sometimes good. This result also indicates that overall, even if FC-DNN cannot find the optimal solution more accurately, its calculated channel and power are better than Dist-Att.

*E. Complexity and Efficiency of Resource Allocation Models*

We have verified the performance of the model from the perspectives of accuracy and CDF. However, during the actual deployment phase, it is crucial to take into account the computational time and required resources. Table IV illustrates the prediction time of the models for the entire test set. The introduction of DL significantly reduces the time consumption compared to exhaustive algorithm, enabling real-time resource allocation.

TABLE IV. CALCULATION TIME (S) FOR TEST SET

| Optimal | FC-DNN | SP-Conv-Att | Dist-Att | Dist-Att-Conv |
|---|---|---|---|---|
| $3.68\times10^4$ | 5.62 | 5.70 | 5.54 | 5.60 |

Further, we use the training parameters of the models to characterize the required computational resources. In this experiment, the parameters mainly come from the weights and biases of modules such as Convolution, Dense, Norm and Attention. In other scenarios, if only the number of users and channels changes, it will only affect the first and last layer of the models. Moreover, the parameters of the intermediate hidden layers need to be determined based on the actual situation. Generally, as the sample size increases and the number of features grows, the required number of hidden layers/neurons is likely to increase. The parameters of the model in this scenario are shown in Table V. Among them, the SP-Conv-Att and the Dist-Att-Conv introduce a large number of parameters due to the convolution operations, the Dist-Att reduces a certain number of parameters through an attention module. Specifically, In Dist-Att-Conv, the parameter burdened by the BS is 1.14M and the calculation of the remaining parameters needs to be afforded by the users. Therefore, if the computational resources of the BS are sufficient, selecting the SP-Conv-Att will achieve better resource allocation performance. If the users have more resources, selecting Dist-Att-Conv will alleviate the burden at the BS. When the resources of both the users and the BS are constrained, selecting Dist-Att, which will further reduce the parameters by combining distribution and attention feature fusion.

TABLE V.  COMPUTATIONAL PARAMETER OF THE MODELS

| FC-DNN | SP-Conv-Att | Dist-Att | Dist-Att-Conv |
|---|---|---|---|
| 2.03M | 5.07M | 0.74M | 4.70M |

V. CONCLUSIONS

This paper first defines an objective function in the context of FD DUEs and HD CUEs, aiming to maximize the SE of DUEs while ensuring that the SE of CUEs reaches a threshold. Then, to better approximate the optimal solution of the objective function, considering the three-dimensional CSI information constituted by users and channels, we propose a centralized model (SP-Conv-Att) and two distributed models (Dist-Att and Dist-Att-Conv) based on spatial convolution and attention mechanisms. These models will accomplish the channel allocation and the power level classification tasks through supervised learning. Simulation results demonstrate that the proposed models are more computationally efficient than the optimal scheme, have higher accuracy and better CDF curves than traditional ERP algorithm and recent DNN model. They will be applicable to different resource allocation scenarios.